\documentclass[aps,prd,amssymb,twocolumn,superscriptaddress,floatfix,nofootinbib,10pt]{revtex4-2}
\usepackage{graphicx}
\usepackage{dcolumn}
\usepackage{hyperref}
\usepackage{amsmath}
\usepackage{amsfonts}
\usepackage{booktabs}
\usepackage{amssymb}
\usepackage[usenames]{color}
\usepackage{lineno}
\usepackage{bm}
\usepackage{mathptmx}
\usepackage{multirow}
\usepackage{braket}
\usepackage{tikz}
\usepackage{url}
\usepackage[compat=1.1.0]{tikz-feynman}

\tikzfeynmanset{arrow size=1}

\begin{document}

\preprint{APS/123-QED}

\title{Scattering observables and correlation function for  $p ~f_1(1285)$ revisited}
 
\author{P. Encarnación}
	\affiliation{Departamento de F\'{i}sica Teórica and IFIC, Centro Mixto Universidad de Valencia-CSIC, Institutos de Investigaci\'{o}n de Paterna, Aptdo. 22085, E-46071 Valencia, Spain}
 
\author{A. Feijoo}
	\affiliation{Instituto de F\'{i}sica Corpuscular, Centro Mixto Universidad de Valencia-CSIC, Institutos de Investigaci\'{o}n de Paterna, Aptdo. 22085, E-46071 Valencia, Spain}
 
\author{E. Oset}
	\affiliation{Departamento de F\'{i}sica Teórica and IFIC, Centro Mixto Universidad de Valencia-CSIC, Institutos de Investigaci\'{o}n de Paterna, Aptdo. 22085, E-46071 Valencia, Spain}
    \affiliation{Department of Physics, Guangxi Normal University, Guilin 541004, China}
 
\date{\today}

\begin{abstract}
In view of the recent theoretical developments in the fixed center approximation for the scattering of a particle with a a two-body cluster, implementing elastic unitarity on the standard fixed center formalism, and the imminent availability of ALICE data on the correlation function of the $p~f_1(1285)$ system, we update the results of a previous work for this correlation function and the low-energy scattering observables. The new results show appreciable changes in some observables and should provide valuable input for comparison with the forthcoming experimental data. Such a comparison is expected to yield relevant information on the nature of the axial-vector meson states.

\end{abstract}

\maketitle


\section{Introduction}
\label{sec:introduction}

Correlation functions (CFs) are increasingly recognized as an important source of information on hadron interactions. By measuring pairs of particles produced in high-energy proton--proton, proton--nucleus, and nucleus--nucleus collisions, and comparing the probability of producing these pairs within the same event with the uncorrelated probability obtained from mixed events, one can extract information that provides access to low-energy scattering observables governing the interaction between particle pairs \cite{Fabbietti:2020bfg,Liu:2024uxn,Albaladejo:2024lam}.

A major advantage of CFs is that they make it possible to investigate particle pairs that are beyond the reach of conventional scattering experiments. While most of the experimental information concerns the interaction of two elementary particles, studies of three-particle CFs have also recently begun \cite{DelGrande:2021mju,ALICE:2022boj,ALICE:2023gxp,ALICE:2023bny,Garrido:2024pwi}. Another line of research has emerged from the measurement of CFs involving a stable particle and a resonance, in particular in studies of the interaction between the proton and the $f_1(1285)$ resonance \cite{Otoninfo,ALICE:2024rjz}. Such studies, not accessible in conventional scattering experiments, open a new avenue to investigate the structure of these resonances, an issue that remains under active debate in hadron physics \cite{Guo:2017jvc,Olsen:2017bmm,Chen:2016qju,Liu:2019zoy,Ali:2017jda,Karliner:2017qhf,Guo:2013sya,Wu:2022ftm,Meng:2022ozq,Liu:2024uxn}.

Anticipating important developments in this line of work, theoretical investigations were initiated in \cite{Encarnacion:2025lyf}, where the $p~f_1(1285)$ CF was evaluated, along with the scattering length and effective range characterizing the interaction of this particle pair. The first step in \cite{Encarnacion:2025lyf} was to assume that the $f_1(1285)$ is not a conventional $q\bar q$ state, but rather a molecular state of $K^*\bar{K}-\bar{K}^*K$ in isospin $I=0$, as supported by numerous independent studies \cite{Lutz:2003fm,Roca:2005nm,Garcia-Recio:2010enl,Zhou:2014ila,Geng:2015yta,Lu:2016nlp}. The second ingredient was the use of the fixed center approximation (FCA) to the Faddeev equations to address the three-body interaction problem. In the FCA, a cluster (here, the $f_1(1285)$) is assumed to remain unchanged during the interaction. This method has been widely applied in similar studies~\cite{Foldy:1945zz,Brueckner:1953zz,Chand:1962ec,Barrett:1999cw,Kamalov:2000iy,MartinezTorres:2020hus,Roca:2010tf}, although it involves approximations that are detailed in Appendix A of Ref.~\cite{Malabarba:2024hlv}. In summary, the FCA is based on Faddeev equations to the three body problem and uses two body scattering amplitudes as input. There are several approximations made, the most important being that there is a cluster of two particles which is not destroyed in the intermediate steps of the reaction. In addition one also assumes that the external momenta of the interacting particles are small compared with those in the loops in the intermediate steps. And a third approximation is that in the propagation of the internal particles, the form factor $F(\vec q_1- \vec q_2)$ is given in average by $F(\vec q_1)F(\vec q_2)$, which allows to obtain a factorized formula.

The work of~\cite{Encarnacion:2025lyf} predicts a CF that deviates significantly from unity, corresponding to a strongly interacting system, as shown by the qualitative studies in \cite{Liu:2023uly}. In fact, in addition to the CF, a bound state with a binding energy of approximately $40$~MeV was also identified in~\cite{Encarnacion:2025lyf}.   

To evaluate the CF and the scattering parameters ($a_0$ and $r_0$), a framework that preserves elastic unitarity at the $p~f_1(1285)$ threshold is required. It was found in~\cite{Encarnacion:2025lyf} that the standard FCA scheme does not fully satisfy this condition. The explicit violation of unitarity in \cite{Encarnacion:2025lyf} could be effectively corrected by multiplying the FCA amplitude by a factor $1.5$. This adjustment provided a temporary fix, allowing for semiquantitative predictions before experimental data could be analysed. However, the underlying problem of unitarity violation remained unresolved, highlighting the need for a more satisfactory solution. This solution was provided in the work of~\cite{Ikeno:2025bsx}. By analogy with particle-nucleus interactions—where an optical potential is defined and then used in the Schrödinger equation to obtain the particle-nucleus scattering matrix \cite{Ericson:1988gk,Seki:1983sh,Nieves:1993ev,Brown:1975di}—the FCA amplitude was identified as the equivalent of an optical potential. In this framework, the extra propagation of the $p$ and the $f_1(1285)$ as a whole is included to restore a unitary amplitude. The new framework was used to study the $n~\bar D_{s0}^*(2317)$ interaction, where the $ D_{s0}^*(2317)$ was considered essentially as a $KD$ bound state, as interpreted in different studies~\cite{vanBeveren:2003kd,Barnes:2003dj,Chen:2004dy,Kolomeitsev:2003ac,Gamermann:2006nm,Guo:2006rp,Yang:2021tvc,Liu:2022dmm}. Once again, a bound state of the $n~\bar D_{s0}^*(2317)$ system was found in~\cite{Ikeno:2025bsx}.

Work was continued along this line, and by employing the formalism of~\cite{Ikeno:2025bsx}, the $n~\bar D_{s1}^*(2460)$ and $n~\bar D_{s1}^*(2536)$ systems were studied, with predictions of states below the corresponding thresholds~\cite{Agatao:2025ckp}. A particularly useful finding of~\cite{Agatao:2025ckp} was that the unitary formula of the scattering matrix of~\cite{Ikeno:2025bsx} could be simplified to a form that allowed an analytical proof of the amplitude’s unitarity.

The new unitary formalism has also been applied to predict an exotic three-body bound state of $K^{*+}D^{*+}K^{*+}$ with total spin $J=3$~\cite{Jia:2025obs}, and more recently to the study of the CF of the $K~f_1(1285)$ system, where a bound state is again predicted \cite{Jia:2026dpl}. In this latter work additional cutoff factors were also implemented to account for off-shell effects.
  
Motivated by these developments, the upcoming experimental data from the ALICE collaboration, and the proximity of available ALICE measurements for the $p\,f_1(1285)$ CF, we find it most appropriate to update the theoretical results of~\cite{Encarnacion:2025lyf} by incorporating the new developments discussed above. The aim of the present work is to provide the most accurate theoretical prediction up to date for the $p\,f_1(1285)$ CF, thereby enabling a meaningful comparison between the forthcoming experimental data and theory.

\section{Formalism}

We follow directly the work of \cite{Jia:2026dpl} in the study of the $Kf_1(1285)$ interaction. The dynamics of this system and that of $pf_1(1285)$ are rather different, since the former involves $KK,K\bar K,KK^*,K \bar K^*$ interactions and the latter $pK,p\bar K, pK^*, p\bar K^*$, which are not related. However, formally they are very similar since both $K$ and $p$ have isospin $1/2$, such that the formalism can be easily mapped from one system to the other.

The $f_1(1285)$ as a function of $K\bar K^*,\bar KK^*$ is given by
$$ f_1(1285)=\frac{1}{2}\big( K^{*+}K^- + K^{*0}\bar K^{0} - K^{*-}K^+ - \bar K^{*0} K^0 \big) $$
which, with the prescription of isospin phases $(K^+,K^0)$, $(\bar K^0,-K^-)$, $(K^{*+},K^{*0})$, $(\bar K^{*0},-K^{*-})$, corresponds to
\begin{equation}
    f_1(1285) = -\frac{1}{\sqrt{2}} (K^*\bar K)^{I=0} - \frac{1}{\sqrt{2}} (\bar K^* K)^{I=0}
    \label{eq:1}
\end{equation}
We shall ignore the sign in Eq.~(\ref{eq:1}), which is irrelevant in the $pf_1\to pf_1$ formalism considered here.

\begin{figure}
    $
    \begin{array}{ccccccc}
        &
        \begin{tikzpicture}[scale=0.9, baseline={(current bounding box.center)}]
            \begin{feynman}
                \vertex (name1) at (0,-0.5) {$(K^*\bar{K})^{I=0}$};
                \vertex (name2) at (0,3.25) {};
                \vertex (iL) at (-0.25,0) {};
                \vertex (iR) at (0.25,0) {};
                \vertex (iN) at (-0.75,0.1) {$N$};
                \vertex (fL) at (-0.25,3) {};
                \vertex (fR) at (0.25,3) {};
                \vertex (fN) at (-0.75,2.9) {$N$};
                \vertex (v1) [dot, style={fill,scale=0.01}] at (-0.25,1.5) {};
                \vertex (v2) [dot, style={fill,scale=0.01}] at (0.25,1.5) {};
                
                \diagram* {
                    (name1) -- [boson, opacity=0.0] (name2),
                    (iN) -- [dashed, fermion] (v1) -- [dashed, fermion] (fN),
                    (iL) -- [fermion] (v1) -- [fermion] (fL),
                    (iR) -- [fermion] (v2) -- [fermion] (fR)
                };
                \filldraw[fill=black!20, draw=black]
                    (0,0.3) ellipse (0.5 and 0.1);
        
                \filldraw[fill=black!20, draw=black]
                    (0,2.7) ellipse (0.5 and 0.1);
                
            \end{feynman}
        \end{tikzpicture}
        & + &
        \begin{tikzpicture}[scale=0.9, baseline={(current bounding box.center)}]
            \begin{feynman}
                \vertex (name1) at (0,-0.5) {$(K^*\bar{K})^{I=0}$};
                \vertex (name2) at (0,3.25) {};
                \vertex (iL) at (-0.25,0) {};
                \vertex (iR) at (0.25,0) {};
                \vertex (iN) at (-0.75,0.1) {$N$};
                \vertex (fL) at (-0.25,3) {};
                \vertex (fR) at (0.25,3) {};
                \vertex (fN) at (0.75,2.9) {$N$};
                \vertex (v1) [dot, style={fill,scale=0.01}] at (-0.25,1.5) {};
                \vertex (v2) [dot, style={fill,scale=0.01}] at (0.25,1.5) {};
                
                \diagram* {
                    (name1) -- [boson, opacity=0.0] (name2),
                    (iN) -- [dashed, fermion] (v1) -- [scalar] (v2) -- [dashed, fermion] (fN),
                    (iL) -- [fermion] (v1) -- [fermion] (fL),
                    (iR) -- [fermion] (v2) -- [fermion] (fR)
                };
                \filldraw[fill=black!20, draw=black]
                    (0,0.3) ellipse (0.5 and 0.1);
        
                \filldraw[fill=black!20, draw=black]
                    (0,2.7) ellipse (0.5 and 0.1);
                
            \end{feynman}
        \end{tikzpicture}
        & + &
        \begin{tikzpicture}[scale=0.9, baseline={(current bounding box.center)}]
            \begin{feynman}
                \vertex (name1) at (0,-0.5) {$(K^*\bar{K})^{I=0}$};
                \vertex (name2) at (0,3.25) {};
                \vertex (iL) at (-0.25,0) {};
                \vertex (iR) at (0.25,0) {};
                \vertex (iN) at (-0.75,0.1) {$N$};
                \vertex (fL) at (-0.25,3) {};
                \vertex (fR) at (0.25,3) {};
                \vertex (fN) at (-0.75,2.9) {$N$};
                \vertex (v1) [dot, style={fill,scale=0.01}] at (-0.25,1) {};
                \vertex (v2) [dot, style={fill,scale=0.01}] at (0.25,1.5) {};
                \vertex (v3) [dot, style={fill,scale=0.01}] at (-0.25,2) {};
                
                \diagram* {
                    (name1) -- [boson, opacity=0.0] (name2),
                    (iN) -- [dashed, fermion] (v1) -- [scalar] (v2) -- [scalar] (v3) -- [dashed, fermion] (fN),
                    (iL) -- [fermion] (v1) -- [fermion] (v3) -- [fermion] (fL),
                    (iR) -- [fermion] (v2) -- [fermion] (fR)
                };
                \filldraw[fill=black!20, draw=black]
                    (0,0.3) ellipse (0.5 and 0.1);
        
                \filldraw[fill=black!20, draw=black]
                    (0,2.7) ellipse (0.5 and 0.1);
                
            \end{feynman}
        \end{tikzpicture}
    & +\ \dots \\
        + &
        \begin{tikzpicture}[scale=0.9, baseline={(current bounding box.center)}]
            \begin{feynman}
                \vertex (name1) at (0,-0.5) {$(K^*\bar{K})^{I=0}$};
                \vertex (name2) at (0,3.25) {};
                \vertex (iL) at (-0.25,0) {};
                \vertex (iR) at (0.25,0) {};
                \vertex (iN) at (0.75,0.1) {$N$};
                \vertex (fL) at (-0.25,3) {};
                \vertex (fR) at (0.25,3) {};
                \vertex (fN) at (0.75,2.9) {$N$};
                \vertex (v1) [dot, style={fill,scale=0.01}] at (-0.25,1.5) {};
                \vertex (v2) [dot, style={fill,scale=0.01}] at (0.25,1.5) {};

                \diagram* {
                    (name1) -- [boson, opacity=0.0] (name2),
                    (iN) -- [dashed, fermion] (v2) -- [dashed, fermion] (fN),
                    (iL) -- [fermion] (v1) -- [fermion] (fL),
                    (iR) -- [fermion] (v2) -- [fermion] (fR)
                };
                \filldraw[fill=black!20, draw=black]
                    (0,0.3) ellipse (0.5 and 0.1);
        
                \filldraw[fill=black!20, draw=black]
                    (0,2.7) ellipse (0.5 and 0.1);
            \end{feynman}
        \end{tikzpicture}
        & + &
        \begin{tikzpicture}[scale=0.9, baseline={(current bounding box.center)}]
            \begin{feynman}
                \vertex (name1) at (0,-0.5) {$(K^*\bar{K})^{I=0}$};
                \vertex (name2) at (0,3.25) {};
                \vertex (iL) at (-0.25,0) {};
                \vertex (iR) at (0.25,0) {};
                \vertex (iN) at (0.75,0.1) {$N$};
                \vertex (fL) at (-0.25,3) {};
                \vertex (fR) at (0.25,3) {};
                \vertex (fN) at (-0.75,2.9) {$N$};
                \vertex (v1) [dot, style={fill,scale=0.01}] at (-0.25,1.5) {};
                \vertex (v2) [dot, style={fill,scale=0.01}] at (0.25,1.5) {};

                \diagram* {
                    (name1) -- [boson, opacity=0.0] (name2),
                    (iN) -- [dashed, fermion] (v2) -- [scalar] (v1) -- [dashed, fermion] (fN),
                    (iL) -- [fermion] (v1) -- [fermion] (fL),
                    (iR) -- [fermion] (v2) -- [fermion] (fR)
                };
                \filldraw[fill=black!20, draw=black]
                    (0,0.3) ellipse (0.5 and 0.1);
        
                \filldraw[fill=black!20, draw=black]
                    (0,2.7) ellipse (0.5 and 0.1);
            \end{feynman}
        \end{tikzpicture}
        & + &
        \begin{tikzpicture}[scale=0.9, baseline={(current bounding box.center)}]
            \begin{feynman}
                \vertex (name1) at (0,-0.5) {$(K^*\bar{K})^{I=0}$};
                \vertex (name2) at (0,3.25) {};
                \vertex (iL) at (-0.25,0) {};
                \vertex (iR) at (0.25,0) {};
                \vertex (iN) at (0.75,0.1) {$N$};
                \vertex (fL) at (-0.25,3) {};
                \vertex (fR) at (0.25,3) {};
                \vertex (fN) at (0.75,2.9) {$N$};
                \vertex (v1) [dot, style={fill,scale=0.01}] at (-0.25,1.5) {};
                \vertex (v2) [dot, style={fill,scale=0.01}] at (0.25,1) {};
                \vertex (v3) [dot, style={fill,scale=0.01}] at (0.25,2) {};

                \diagram* {
                    (name1) -- [boson, opacity=0.0] (name2),
                    (iN) -- [dashed, fermion] (v2) -- [scalar] (v1) -- [scalar] (v3) -- [dashed, fermion] (fN),
                    (iL) -- [fermion] (v1) -- [fermion] (fL),
                    (iR) -- [fermion] (v2) -- [fermion] (v3) -- [fermion] (fR)
                };
                \filldraw[fill=black!20, draw=black]
                    (0,0.3) ellipse (0.5 and 0.1);
        
                \filldraw[fill=black!20, draw=black]
                    (0,2.7) ellipse (0.5 and 0.1);
            \end{feynman}
        \end{tikzpicture}
    & +\ \dots
    \end{array}
    $
    \caption{Diagrams in the FCA for $Nf_1(1285)$ scattering. The diagrams for $(\bar K ^* K)^{I=0}$ have to be considered in addition.}
    \label{fig:diagrams}
\end{figure}
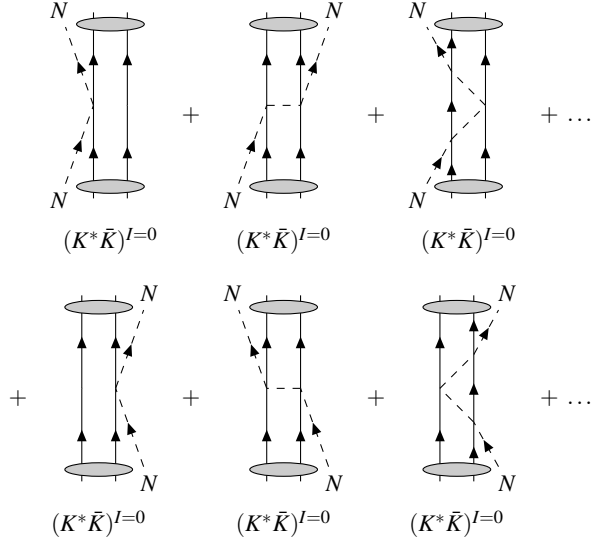

The FCA sums the diagrams depicted in Fig.~\ref{fig:diagrams}. The $K^* \bar K$ and $\bar K^* K$ systems are not mixed in this approach because $pK^*$ cannot transition to $p\bar K^*$. One then evaluates the $p K^* \bar K \to p K^* \bar K$ and $p\bar K^* K \to p \bar K^* K$ transitions independently and takes the average at the end.

By taking into account the $I=0$ character of the $K^* \bar K$ cluster, the $pK^*$ amplitude, $t_1$, and the $p\bar K$ amplitude, $t_2$, can be written as
\begin{eqnarray}
    t_1 &=& \frac{3}{4} t_{pK^*}^{(1)} + \frac{1}{4} t_{pK^*}^{(0)} \nonumber \\
    t_2 &=& \frac{3}{4} t_{p\bar{K}}^{(1)} + \frac{1}{4} t_{p\bar{K}}^{(0)} 
\end{eqnarray}
where the superscript denotes the isospin component of the interaction. Similarly, considering the $I=0$ of the $K^*\bar K$ system, one also writes
\begin{eqnarray}
\label{t_prime}
    t_1' &=& \frac{3}{4} t_{p\bar{K}^*}^{(1)} + \frac{1}{4} t_{p\bar{K}^*}^{(0)} \nonumber \\
    t_2' &=& \frac{3}{4} t_{pK}^{(1)} + \frac{1}{4} t_{pK}^{(0)} 
\end{eqnarray}

We will refer the final amplitude for the three-body system to the normalization of the $p$ and $f_1$ fields in the Mandl and Shaw normalization \cite{mandl_shaw_2010} and then it is convenient to use the differently normalized amplitude
\begin{eqnarray}
    \widetilde t_1 = \frac{M_C}{M_{K^*}}t_1 \quad;\quad\widetilde t_2 = \frac{M_C}{M_K}t_2
\end{eqnarray}
and similarly for $t_1',t_2'$, where $M_C$ is the mass of the cluster $f_1(1285)$. These amplitudes are evaluated in the Appendix of \cite{Encarnacion:2025lyf}.

These amplitudes depend on the variables $s_i$ given by
\begin{eqnarray}
    s_1(NK^*)&=&m_N^2+(\xi m_{K^*})^2 + 2\xi m_{K^*} q^0 \nonumber \\
    s_2(N\bar{K})&=&m_N^2+(\xi m_{\bar{K}})^2 + 2\xi m_{\bar{K}} q^0
    \label{eq:5}
\end{eqnarray}
with $\xi = M_C/(m_{K^*}+m_K)$ and $q_0$, the proton momentum in the cluster rest frame, given by
\begin{equation}
    q^0=\frac{s-M_N^2-M_C^2}{2M_c},
\end{equation}
where $M_N$ is the mass of the proton. Eq.~(\ref{eq:5}) assumes that the binding energy of the $f_1(1285)$ is shared between the $K^*$ and $\bar K$ proportionally to their mass.

For Fig.~(\ref{fig:diagrams}) it is useful to put together the diagrams where the interaction of the proton begins from particle $i$ ($i=1$ for $K^*$ and $i=2$ for $\bar K$) and finishes in particle $j$, which we call $\widetilde T_{ij}$ and which are given by
\begin{eqnarray}
    && \widetilde T_{11} = \frac{\widetilde t_1}{1-\widetilde t_1\widetilde t_2 G_0^2} \quad ; \quad \widetilde T_{22} = \frac{\widetilde t_2}{1-\widetilde t_1\widetilde t_2 G_0^2} \nonumber \\
    &&\widetilde T_{12} = \widetilde T_{21} = \frac{\widetilde t_1 \widetilde t_2 G_0}{1-\widetilde t_1\widetilde t_2 G_0^2} 
\end{eqnarray}
where $G_0$ denotes the propagator of the proton from particle 1 to particle 2, folded by the cluster wave function, given by
\begin{eqnarray}
    G_0(\sqrt{s}) = \int &&\frac{d^3 q}{(2\pi)^3} \frac{2M_N }{2E(q)2\omega_C(q)} \frac{F_C(q)}{\sqrt{s}-E(q)-\omega_C(q)+i\epsilon} \nonumber \\
    &&\times \theta(q_{max}^{(1)}-q_1^*)\theta(q_{max}^{(2)}-q_2^*)
    \label{eq:8}
\end{eqnarray}
where $\omega_C(q)=\sqrt{M_C^2+q^2}$, $E(q)=\sqrt{M_N^2+q^2}$, with $q^{(1)}_{max},q^{(2)}_{max}$ the regulators of the $pK^*, p \bar K$ loop functions in the cutoff regularization of the amplitudes (see Appendix of \cite{Encarnacion:2025lyf}), and $q_1^*,q_2^*$ the center-of-mass momenta of the proton in the $pK^*,p\bar K$ rest frame, for the intermediate $p$, assuming the external proton momentum to be zero, given by
\begin{eqnarray}
    \vec q_i^* = \vec q \left( 1 - \frac{1}{2} \frac{M_N}{M_N+M_i} \right)
\end{eqnarray}
with $M_1 = M_{K^*}$, $M_2 = M_{\bar K}$.

In Eq.~(\ref{eq:8}) one has also $F_C(q)$, which is the cluster form factor, given by \cite{Roca:2010tf}
\begin{equation}
    F_c(q)=\tilde{F}_c(q)/\tilde{F}_c(0)
\end{equation}
with
\begin{eqnarray}
    \tilde{F}_c(q) = &\displaystyle\int&\frac{d^3p}{(2\pi)^3}\frac{1}{M_c-\omega_{K^*}(p)-\omega_{\bar{K}}(p)} \nonumber \\
    &\times& \frac{1}{M_c-\omega_{K^*}(\vec{p}-\vec{q})-\omega_{\bar{K}}(\vec{p}-\vec{q})} \nonumber \\
    &|\vec{p}|&<q_{max} \nonumber \\
    &|\vec{p}-\vec{q}|&<q_{max}
    \label{form_factor}
\end{eqnarray}
where $q_{max}$ is the cutoff momentum needed to regularize the loops in the chiral unitary approach used to generate $f_1(1285)$, chosen to be $q_{max}=1000$ MeV, ensuring that the $f_1(1285)$ state is produced at its experimental mass.

For the following it is useful to write $\widetilde T_{ij}$ in matrix form,
\begin{eqnarray}
    \widetilde T = 
    \begin{pmatrix}
        \widetilde T_{11} & \widetilde T_{12} \\
        \widetilde T_{21} & \widetilde T_{22}
    \end{pmatrix}.
\end{eqnarray}

\subsection{Elastic unitarization}

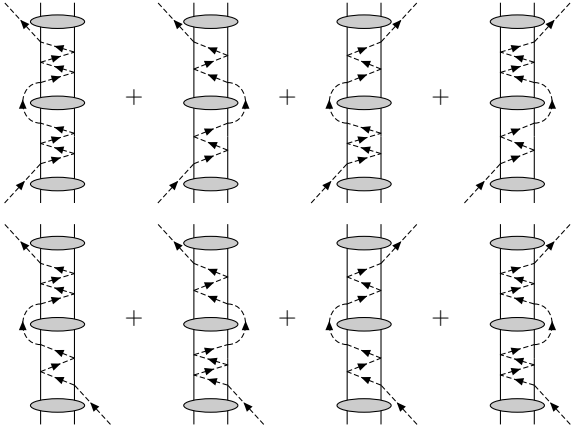
\begin{figure}
    \centering
        \begin{tikzpicture}[scale=0.9, baseline={(current bounding box.center)}, inner sep=0pt, outer sep=0pt]]
            \begin{feynman}
                \vertex (iL) at (-0.25,0) {};
                \vertex (iR) at (0.25,0) {};
                \vertex (iN) at (-0.8,0) {};
                \vertex (fL) at (-0.25,3) {};
                \vertex (fR) at (0.25,3) {};
                \vertex (fN) at (-0.8,3) {};
                \vertex (ghost1) at (-0.8,0) {};
                \vertex (ghost2) at (0.8,3) {};
                
                \vertex (v11) [dot, style={fill,scale=0.01}] at (-0.25,0.6) {};
                \vertex (v12) [dot, style={fill,scale=0.01}] at (-0.25,1.2) {};
                
                \vertex (v21) [dot, style={fill,scale=0.01}] at (-0.25,1.8) {};
                \vertex (v22) [dot, style={fill,scale=0.01}] at (-0.25,2.4) {};
                
                \vertex (v111) [dot, style={fill,scale=0.01}] at (0.25,0.75) {};
                \vertex (v112) [dot, style={fill,scale=0.01}] at (-0.25,0.9) {};
                \vertex (v113) [dot, style={fill,scale=0.01}] at (0.25,1.05) {};
                
                \vertex (v211) [dot, style={fill,scale=0.01}] at (0.25,1.95) {};
                \vertex (v212) [dot, style={fill,scale=0.01}] at (-0.25,2.1) {};
                \vertex (v213) [dot, style={fill,scale=0.01}] at (0.25,2.25) {};
                
                \diagram* {
                    (iN) -- [dashed, fermion, arrow size=0.8pt, dash pattern=on 2pt off 1pt] (v11) 
                    -- [dashed, fermion, arrow size=0.8pt, dash pattern=on 2pt off 1pt] (v111) 
                    -- [dashed, fermion, arrow size=0.8pt, dash pattern=on 2pt off 1pt] (v112) 
                    -- [dashed, fermion, arrow size=0.8pt, dash pattern=on 2pt off 1pt] (v113) 
                    -- [dashed, fermion, arrow size=0.8pt, dash pattern=on 2pt off 1pt] (v12) 
                    -- [dashed, fermion, arrow size=0.8pt, dash pattern=on 2pt off 1pt, half left] (v21) 
                    -- [dashed, fermion, arrow size=0.8pt, dash pattern=on 2pt off 1pt] (v211) 
                    -- [dashed, fermion, arrow size=0.8pt, dash pattern=on 2pt off 1pt] (v212) 
                    -- [dashed, fermion, arrow size=0.8pt, dash pattern=on 2pt off 1pt] (v213) 
                    -- [dashed, fermion, arrow size=0.8pt, dash pattern=on 2pt off 1pt] (v22) 
                    -- [dashed, fermion, arrow size=0.8pt, dash pattern=on 2pt off 1pt] (fN) ,
                    (iL) -- (v1) --  (fL),
                    (iR) --  (v2) --  (fR),
                    (ghost1) -- [boson, opacity=0.0] (ghost2),
                };
                \filldraw[fill=black!20, draw=black]
                    (0,0.3) ellipse (0.4 and 0.1);
                \filldraw[fill=black!20, draw=black]
                    (0,1.5) ellipse (0.4 and 0.1);
                \filldraw[fill=black!20, draw=black]
                    (0,2.7) ellipse (0.4 and 0.1);
                
            \end{feynman}
        \end{tikzpicture} 
        \hspace{0pt} $+$ \hspace{0pt} 
        \begin{tikzpicture}[scale=0.9, baseline={(current bounding box.center)}, inner sep=0pt, outer sep=0pt]]
            \begin{feynman}
                \vertex (iL) at (-0.25,0) {};
                \vertex (iR) at (0.25,0) {};
                \vertex (iN) at (-0.8,0) {};
                \vertex (fL) at (-0.25,3) {};
                \vertex (fR) at (0.25,3) {};
                \vertex (fN) at (-0.8,3) {};
                \vertex (ghost1) at (-0.8,0) {};
                \vertex (ghost2) at (0.8,3) {};
                
                \vertex (v11) [dot, style={fill,scale=0.01}] at (-0.25,0.6) {};
                \vertex (v12) [dot, style={fill,scale=0.01}] at (0.25,1.2) {};
                
                \vertex (v21) [dot, style={fill,scale=0.01}] at (0.25,1.8) {};
                \vertex (v22) [dot, style={fill,scale=0.01}] at (-0.25,2.4) {};
                
                \vertex (v111) [dot, style={fill,scale=0.01}] at (0.25,0.8) {};
                \vertex (v112) [dot, style={fill,scale=0.01}] at (-0.25,1) {};
                
                \vertex (v211) [dot, style={fill,scale=0.01}] at (-0.25,2) {};
                \vertex (v212) [dot, style={fill,scale=0.01}] at (0.25,2.2) {};
                
                \diagram* {
                    (iN) -- [dashed, fermion, arrow size=0.8pt, dash pattern=on 2pt off 1pt] (v11) 
                    -- [dashed, fermion, arrow size=0.8pt, dash pattern=on 2pt off 1pt] (v111) 
                    -- [dashed, fermion, arrow size=0.8pt, dash pattern=on 2pt off 1pt] (v112) 
                    -- [dashed, fermion, arrow size=0.8pt, dash pattern=on 2pt off 1pt] (v12) 
                    -- [dashed, fermion, arrow size=0.8pt, dash pattern=on 2pt off 1pt, half right] (v21) 
                    -- [dashed, fermion, arrow size=0.8pt, dash pattern=on 2pt off 1pt] (v211) 
                    -- [dashed, fermion, arrow size=0.8pt, dash pattern=on 2pt off 1pt] (v212) 
                    -- [dashed, fermion, arrow size=0.8pt, dash pattern=on 2pt off 1pt] (v22) 
                    -- [dashed, fermion, arrow size=0.8pt, dash pattern=on 2pt off 1pt] (fN) ,
                    (iL) -- (v1) --  (fL),
                    (iR) --  (v2) --  (fR),
                    (ghost1) -- [boson, opacity=0.0] (ghost2),
                };
                \filldraw[fill=black!20, draw=black]
                    (0,0.3) ellipse (0.4 and 0.1);
                \filldraw[fill=black!20, draw=black]
                    (0,1.5) ellipse (0.4 and 0.1);
                \filldraw[fill=black!20, draw=black]
                    (0,2.7) ellipse (0.4 and 0.1);
                
            \end{feynman}
        \end{tikzpicture}
        \hspace{0pt} $+$ \hspace{0pt} 
        \begin{tikzpicture}[scale=0.9, baseline={(current bounding box.center)}, inner sep=0pt, outer sep=0pt]]
            \begin{feynman}
                \vertex (iL) at (-0.25,0) {};
                \vertex (iR) at (0.25,0) {};
                \vertex (iN) at (-0.8,0) {};
                \vertex (fL) at (-0.25,3) {};
                \vertex (fR) at (0.25,3) {};
                \vertex (fN) at (0.8,3) {};
                \vertex (ghost1) at (-0.8,0) {};
                \vertex (ghost2) at (0.8,3) {};
                
                \vertex (v11) [dot, style={fill,scale=0.01}] at (-0.25,0.6) {};
                \vertex (v12) [dot, style={fill,scale=0.01}] at (-0.25,1.2) {};
                
                \vertex (v21) [dot, style={fill,scale=0.01}] at (-0.25,1.8) {};
                \vertex (v22) [dot, style={fill,scale=0.01}] at (0.25,2.4) {};
                
                \vertex (v111) [dot, style={fill,scale=0.01}] at (0.25,0.75) {};
                \vertex (v112) [dot, style={fill,scale=0.01}] at (-0.25,0.9) {};
                \vertex (v113) [dot, style={fill,scale=0.01}] at (0.25,1.05) {};
                
                \vertex (v211) [dot, style={fill,scale=0.01}] at (0.25,2) {};
                \vertex (v212) [dot, style={fill,scale=0.01}] at (-0.25,2.2) {};
                
                \diagram* {
                    (iN) -- [dashed, fermion, arrow size=0.8pt, dash pattern=on 2pt off 1pt] (v11) 
                    -- [dashed, fermion, arrow size=0.8pt, dash pattern=on 2pt off 1pt] (v111) 
                    -- [dashed, fermion, arrow size=0.8pt, dash pattern=on 2pt off 1pt] (v112) 
                    -- [dashed, fermion, arrow size=0.8pt, dash pattern=on 2pt off 1pt] (v113) 
                    -- [dashed, fermion, arrow size=0.8pt, dash pattern=on 2pt off 1pt] (v12) 
                    -- [dashed, fermion, arrow size=0.8pt, dash pattern=on 2pt off 1pt, half left] (v21) 
                    -- [dashed, fermion, arrow size=0.8pt, dash pattern=on 2pt off 1pt] (v211) 
                    -- [dashed, fermion, arrow size=0.8pt, dash pattern=on 2pt off 1pt] (v212) 
                    -- [dashed, fermion, arrow size=0.8pt, dash pattern=on 2pt off 1pt] (v22) 
                    -- [dashed, fermion, arrow size=0.8pt, dash pattern=on 2pt off 1pt] (fN) ,
                    (iL) -- (v1) --  (fL),
                    (iR) --  (v2) --  (fR),
                    (ghost1) -- [boson, opacity=0.0] (ghost2),
                };
                \filldraw[fill=black!20, draw=black]
                    (0,0.3) ellipse (0.4 and 0.1);
                \filldraw[fill=black!20, draw=black]
                    (0,1.5) ellipse (0.4 and 0.1);
                \filldraw[fill=black!20, draw=black]
                    (0,2.7) ellipse (0.4 and 0.1);
                
            \end{feynman}
        \end{tikzpicture}
        \hspace{0pt} $+$ \hspace{0pt} 
        \begin{tikzpicture}[scale=0.9, baseline={(current bounding box.center)}, inner sep=0pt, outer sep=0pt]]
            \begin{feynman}
                \vertex (iL) at (-0.25,0) {};
                \vertex (iR) at (0.25,0) {};
                \vertex (iN) at (-0.8,0) {};
                \vertex (fL) at (-0.25,3) {};
                \vertex (fR) at (0.25,3) {};
                \vertex (fN) at (0.8,3) {};
                \vertex (ghost1) at (-0.8,0) {};
                \vertex (ghost2) at (0.8,3) {};
                
                \vertex (v11) [dot, style={fill,scale=0.01}] at (-0.25,0.6) {};
                \vertex (v12) [dot, style={fill,scale=0.01}] at (0.25,1.2) {};
                
                \vertex (v21) [dot, style={fill,scale=0.01}] at (0.25,1.8) {};
                \vertex (v22) [dot, style={fill,scale=0.01}] at (0.25,2.4) {};
                
                \vertex (v111) [dot, style={fill,scale=0.01}] at (0.25,0.8) {};
                \vertex (v112) [dot, style={fill,scale=0.01}] at (-0.25,1) {};
                
                \vertex (v211) [dot, style={fill,scale=0.01}] at (-0.25,1.95) {};
                \vertex (v212) [dot, style={fill,scale=0.01}] at (0.25,2.1) {};
                \vertex (v213) [dot, style={fill,scale=0.01}] at (-0.25,2.25) {};
                
                \diagram* {
                    (iN) -- [dashed, fermion, arrow size=0.8pt, dash pattern=on 2pt off 1pt] (v11) 
                    -- [dashed, fermion, arrow size=0.8pt, dash pattern=on 2pt off 1pt] (v111) 
                    -- [dashed, fermion, arrow size=0.8pt, dash pattern=on 2pt off 1pt] (v112) 
                    -- [dashed, fermion, arrow size=0.8pt, dash pattern=on 2pt off 1pt] (v12) 
                    -- [dashed, fermion, arrow size=0.8pt, dash pattern=on 2pt off 1pt, half right] (v21) 
                    -- [dashed, fermion, arrow size=0.8pt, dash pattern=on 2pt off 1pt] (v211) 
                    -- [dashed, fermion, arrow size=0.8pt, dash pattern=on 2pt off 1pt] (v212) 
                    -- [dashed, fermion, arrow size=0.8pt, dash pattern=on 2pt off 1pt] (v213) 
                    -- [dashed, fermion, arrow size=0.8pt, dash pattern=on 2pt off 1pt] (v22) 
                    -- [dashed, fermion, arrow size=0.8pt, dash pattern=on 2pt off 1pt] (fN) ,
                    (iL) -- (v1) --  (fL),
                    (iR) --  (v2) --  (fR),
                    (ghost1) -- [boson, opacity=0.0] (ghost2),
                };
                \filldraw[fill=black!20, draw=black]
                    (0,0.3) ellipse (0.4 and 0.1);
                \filldraw[fill=black!20, draw=black]
                    (0,1.5) ellipse (0.4 and 0.1);
                \filldraw[fill=black!20, draw=black]
                    (0,2.7) ellipse (0.4 and 0.1);
                
            \end{feynman}
        \end{tikzpicture} 
        \\ [5pt]
        \begin{tikzpicture}[scale=0.9, baseline={(current bounding box.center)}, inner sep=0pt, outer sep=0pt]]
            \begin{feynman}
                \vertex (iL) at (-0.25,0) {};
                \vertex (iR) at (0.25,0) {};
                \vertex (iN) at (0.8,0) {};
                \vertex (fL) at (-0.25,3) {};
                \vertex (fR) at (0.25,3) {};
                \vertex (fN) at (-0.8,3) {};
                \vertex (ghost1) at (-0.8,0) {};
                \vertex (ghost2) at (0.8,3) {};
                
                \vertex (v11) [dot, style={fill,scale=0.01}] at (0.25,0.6) {};
                \vertex (v12) [dot, style={fill,scale=0.01}] at (-0.25,1.2) {};
                
                \vertex (v21) [dot, style={fill,scale=0.01}] at (-0.25,1.8) {};
                \vertex (v22) [dot, style={fill,scale=0.01}] at (-0.25,2.4) {};
                
                \vertex (v111) [dot, style={fill,scale=0.01}] at (-0.25,0.8) {};
                \vertex (v112) [dot, style={fill,scale=0.01}] at (0.25,1) {};
                
                \vertex (v211) [dot, style={fill,scale=0.01}] at (0.25,1.95) {};
                \vertex (v212) [dot, style={fill,scale=0.01}] at (-0.25,2.1) {};
                \vertex (v213) [dot, style={fill,scale=0.01}] at (0.25,2.25) {};
                
                \diagram* {
                    (iN) -- [dashed, fermion, arrow size=0.8pt, dash pattern=on 2pt off 1pt] (v11) 
                    -- [dashed, fermion, arrow size=0.8pt, dash pattern=on 2pt off 1pt] (v111) 
                    -- [dashed, fermion, arrow size=0.8pt, dash pattern=on 2pt off 1pt] (v112) 
                    -- [dashed, fermion, arrow size=0.8pt, dash pattern=on 2pt off 1pt] (v12) 
                    -- [dashed, fermion, arrow size=0.8pt, dash pattern=on 2pt off 1pt, half left] (v21) 
                    -- [dashed, fermion, arrow size=0.8pt, dash pattern=on 2pt off 1pt] (v211) 
                    -- [dashed, fermion, arrow size=0.8pt, dash pattern=on 2pt off 1pt] (v212) 
                    -- [dashed, fermion, arrow size=0.8pt, dash pattern=on 2pt off 1pt] (v213) 
                    -- [dashed, fermion, arrow size=0.8pt, dash pattern=on 2pt off 1pt] (v22) 
                    -- [dashed, fermion, arrow size=0.8pt, dash pattern=on 2pt off 1pt] (fN) ,
                    (iL) -- (v1) --  (fL),
                    (iR) --  (v2) --  (fR),
                    (ghost1) -- [boson, opacity=0.0] (ghost2),
                };
                \filldraw[fill=black!20, draw=black]
                    (0,0.3) ellipse (0.4 and 0.1);
                \filldraw[fill=black!20, draw=black]
                    (0,1.5) ellipse (0.4 and 0.1);
                \filldraw[fill=black!20, draw=black]
                    (0,2.7) ellipse (0.4 and 0.1);
                
            \end{feynman}
        \end{tikzpicture}
        \hspace{0pt} $+$ \hspace{0pt} 
        \begin{tikzpicture}[scale=0.9, baseline={(current bounding box.center)}, inner sep=0pt, outer sep=0pt]]
            \begin{feynman}
                \vertex (iL) at (-0.25,0) {};
                \vertex (iR) at (0.25,0) {};
                \vertex (iN) at (0.8,0) {};
                \vertex (fL) at (-0.25,3) {};
                \vertex (fR) at (0.25,3) {};
                \vertex (fN) at (-0.8,3) {};
                \vertex (ghost1) at (-0.8,0) {};
                \vertex (ghost2) at (0.8,3) {};
                
                \vertex (v11) [dot, style={fill,scale=0.01}] at (0.25,0.6) {};
                \vertex (v12) [dot, style={fill,scale=0.01}] at (0.25,1.2) {};
                
                \vertex (v21) [dot, style={fill,scale=0.01}] at (0.25,1.8) {};
                \vertex (v22) [dot, style={fill,scale=0.01}] at (-0.25,2.4) {};
                
                \vertex (v111) [dot, style={fill,scale=0.01}] at (-0.25,0.75) {};
                \vertex (v112) [dot, style={fill,scale=0.01}] at (0.25,0.9) {};
                \vertex (v113) [dot, style={fill,scale=0.01}] at (-0.25,1.05) {};
                
                \vertex (v211) [dot, style={fill,scale=0.01}] at (-0.25,2) {};
                \vertex (v212) [dot, style={fill,scale=0.01}] at (0.25,2.2) {};
                
                \diagram* {
                    (iN) -- [dashed, fermion, arrow size=0.8pt, dash pattern=on 2pt off 1pt] (v11) 
                    -- [dashed, fermion, arrow size=0.8pt, dash pattern=on 2pt off 1pt] (v111) 
                    -- [dashed, fermion, arrow size=0.8pt, dash pattern=on 2pt off 1pt] (v112) 
                    -- [dashed, fermion, arrow size=0.8pt, dash pattern=on 2pt off 1pt] (v113) 
                    -- [dashed, fermion, arrow size=0.8pt, dash pattern=on 2pt off 1pt] (v12) 
                    -- [dashed, fermion, arrow size=0.8pt, dash pattern=on 2pt off 1pt, half right] (v21) 
                    -- [dashed, fermion, arrow size=0.8pt, dash pattern=on 2pt off 1pt] (v211) 
                    -- [dashed, fermion, arrow size=0.8pt, dash pattern=on 2pt off 1pt] (v212) 
                    -- [dashed, fermion, arrow size=0.8pt, dash pattern=on 2pt off 1pt] (v22) 
                    -- [dashed, fermion, arrow size=0.8pt, dash pattern=on 2pt off 1pt] (fN) ,
                    (iL) -- (v1) --  (fL),
                    (iR) --  (v2) --  (fR),
                    (ghost1) -- [boson, opacity=0.0] (ghost2),
                };
                \filldraw[fill=black!20, draw=black]
                    (0,0.3) ellipse (0.4 and 0.1);
                \filldraw[fill=black!20, draw=black]
                    (0,1.5) ellipse (0.4 and 0.1);
                \filldraw[fill=black!20, draw=black]
                    (0,2.7) ellipse (0.4 and 0.1);
                
            \end{feynman}
        \end{tikzpicture}
        \hspace{0pt} $+$ \hspace{0pt} 
        \begin{tikzpicture}[scale=0.9, baseline={(current bounding box.center)}, inner sep=0pt, outer sep=0pt]]
            \begin{feynman}
                \vertex (iL) at (-0.25,0) {};
                \vertex (iR) at (0.25,0) {};
                \vertex (iN) at (0.8,0) {};
                \vertex (fL) at (-0.25,3) {};
                \vertex (fR) at (0.25,3) {};
                \vertex (fN) at (0.8,3) {};
                \vertex (ghost1) at (-0.8,0) {};
                \vertex (ghost2) at (0.8,3) {};
                
                \vertex (v11) [dot, style={fill,scale=0.01}] at (0.25,0.6) {};
                \vertex (v12) [dot, style={fill,scale=0.01}] at (-0.25,1.2) {};
                
                \vertex (v21) [dot, style={fill,scale=0.01}] at (-0.25,1.8) {};
                \vertex (v22) [dot, style={fill,scale=0.01}] at (0.25,2.4) {};
                
                \vertex (v111) [dot, style={fill,scale=0.01}] at (-0.25,0.8) {};
                \vertex (v112) [dot, style={fill,scale=0.01}] at (0.25,1) {};
                
                \vertex (v211) [dot, style={fill,scale=0.01}] at (0.25,2) {};
                \vertex (v212) [dot, style={fill,scale=0.01}] at (-0.25,2.2) {};
                
                \diagram* {
                    (iN) -- [dashed, fermion, arrow size=0.8pt, dash pattern=on 2pt off 1pt] (v11) 
                    -- [dashed, fermion, arrow size=0.8pt, dash pattern=on 2pt off 1pt] (v111) 
                    -- [dashed, fermion, arrow size=0.8pt, dash pattern=on 2pt off 1pt] (v112) 
                    -- [dashed, fermion, arrow size=0.8pt, dash pattern=on 2pt off 1pt] (v12) 
                    -- [dashed, fermion, arrow size=0.8pt, dash pattern=on 2pt off 1pt, half left] (v21) 
                    -- [dashed, fermion, arrow size=0.8pt, dash pattern=on 2pt off 1pt] (v211) 
                    -- [dashed, fermion, arrow size=0.8pt, dash pattern=on 2pt off 1pt] (v212) 
                    -- [dashed, fermion, arrow size=0.8pt, dash pattern=on 2pt off 1pt] (v22) 
                    -- [dashed, fermion, arrow size=0.8pt, dash pattern=on 2pt off 1pt] (fN) ,
                    (iL) -- (v1) --  (fL),
                    (iR) --  (v2) --  (fR),
                    (ghost1) -- [boson, opacity=0.0] (ghost2),
                };
                \filldraw[fill=black!20, draw=black]
                    (0,0.3) ellipse (0.4 and 0.1);
                \filldraw[fill=black!20, draw=black]
                    (0,1.5) ellipse (0.4 and 0.1);
                \filldraw[fill=black!20, draw=black]
                    (0,2.7) ellipse (0.4 and 0.1);
                
            \end{feynman}
        \end{tikzpicture}
        \hspace{0pt} $+$ \hspace{0pt} 
        \begin{tikzpicture}[scale=0.9, baseline={(current bounding box.center)}, inner sep=0pt, outer sep=0pt]
            \begin{feynman}
                \vertex (iL) at (-0.25,0) {};
                \vertex (iR) at (0.25,0) {};
                \vertex (iN) at (0.8,0) {};
                \vertex (fL) at (-0.25,3) {};
                \vertex (fR) at (0.25,3) {};
                \vertex (fN) at (0.8,3) {};
                \vertex (ghost1) at (-0.8,0) {};
                \vertex (ghost2) at (0.8,3) {};
                
                \vertex (v11) [dot, style={fill,scale=0.01}] at (0.25,0.6) {};
                \vertex (v12) [dot, style={fill,scale=0.01}] at (0.25,1.2) {};
                
                \vertex (v21) [dot, style={fill,scale=0.01}] at (0.25,1.8) {};
                \vertex (v22) [dot, style={fill,scale=0.01}] at (0.25,2.4) {};
                
                \vertex (v111) [dot, style={fill,scale=0.01}] at (-0.25,0.75) {};
                \vertex (v112) [dot, style={fill,scale=0.01}] at (0.25,0.9) {};
                \vertex (v113) [dot, style={fill,scale=0.01}] at (-0.25,1.05) {};
                
                \vertex (v211) [dot, style={fill,scale=0.01}] at (-0.25,1.95) {};
                \vertex (v212) [dot, style={fill,scale=0.01}] at (0.25,2.1) {};
                \vertex (v213) [dot, style={fill,scale=0.01}] at (-0.25,2.25) {};
                
                \diagram* {
                    (iN) -- [dashed, fermion, arrow size=0.8pt, dash pattern=on 2pt off 1pt] (v11) 
                    -- [dashed, fermion, arrow size=0.8pt, dash pattern=on 2pt off 1pt] (v111) 
                    -- [dashed, fermion, arrow size=0.8pt, dash pattern=on 2pt off 1pt] (v112) 
                    -- [dashed, fermion, arrow size=0.8pt, dash pattern=on 2pt off 1pt] (v113) 
                    -- [dashed, fermion, arrow size=0.8pt, dash pattern=on 2pt off 1pt] (v12) 
                    -- [dashed, fermion, arrow size=0.8pt, dash pattern=on 2pt off 1pt, half right] (v21) 
                    -- [dashed, fermion, arrow size=0.8pt, dash pattern=on 2pt off 1pt] (v211) 
                    -- [dashed, fermion, arrow size=0.8pt, dash pattern=on 2pt off 1pt] (v212) 
                    -- [dashed, fermion, arrow size=0.8pt, dash pattern=on 2pt off 1pt] (v213) 
                    -- [dashed, fermion, arrow size=0.8pt, dash pattern=on 2pt off 1pt] (v22) 
                    -- [dashed, fermion, arrow size=0.8pt, dash pattern=on 2pt off 1pt] (fN) ,
                    (iL) -- (v1) --  (fL),
                    (iR) --  (v2) --  (fR),
                    (ghost1) -- [boson, opacity=0.0] (ghost2),
                };
                \filldraw[fill=black!20, draw=black]
                    (0,0.3) ellipse (0.4 and 0.1);
                \filldraw[fill=black!20, draw=black]
                    (0,1.5) ellipse (0.4 and 0.1);
                \filldraw[fill=black!20, draw=black]
                    (0,2.7) ellipse (0.4 and 0.1);
                
            \end{feynman}
        \end{tikzpicture}
    \caption{Diagrams required for the unitarization of the $pf_1(1285)$ scattering amplitude.}
    \label{fig:diagramsunitarity}
\end{figure}

Elastic unitarity in $pf_1(1285)$ requires the sum of the diagrams of Fig.~\ref{fig:diagramsunitarity}, and the iteration of these diagrams. The sum of all of them gives rise to the amplitude $\widetilde T_{ij}'$, with the same meaning for the $i,j$ indices as $\widetilde T_{ij}$, which is given in matrix form by
\begin{eqnarray}
    \widetilde T'=[1-\widetilde T G_C ]^{-1} \widetilde T\, ,
\end{eqnarray}
with
\begin{eqnarray}
    G_C = 
    \begin{pmatrix}
        G_C^{(1)} & 0 \\ 0 & G_C^{(2)}\, .
    \end{pmatrix}
\end{eqnarray}
Here,
\begin{eqnarray}
    G_C^{(i)}(\sqrt{s}) = \int && \frac{d^3 q}{(2\pi)^3} \frac{2M_N }{2E(q)2\omega_C(q)} \frac{[F_C^{(i)}(q)]^2}{\sqrt{s}-E(q)-\omega_C(q)+i\epsilon} \nonumber \\
    &&\times \theta(q_{max}^{(i)}-q_i^*)\, ,
\end{eqnarray}
with \cite{Yamagata-Sekihara:2010kpd}
\begin{subequations}
\begin{align}
    &F_C^{(1)}(q) = F_C\left( \frac{m_{\bar K}}{m_{K^*} + m_{\bar K}} q \right) \\
    &F_C^{(2)}(q) = F_C\left( \frac{m_{K^*}}{m_{K^*} + m_{\bar K}} q \right)\, .
\end{align}
\end{subequations}
The total amplitude is derived in \cite{Ikeno:2025bsx}, and in a simplified form in \cite{Agatao:2025ckp}, and reads 
\begin{equation}
    T^{\rm tot}_{K^*\bar K}(\sqrt{s}) = \sum_{i,j} \widetilde T_{ij}' = \frac{\widetilde t_1+\widetilde t_2 + (2G_0-G_C^{(1)}-G_C^{(2)})\widetilde t_1\widetilde t_2}{1-G_C^{(1)}\widetilde t_1-G_C^{(2)}\widetilde t_2-(G_0^2-G_C^{(1)}G_C^{(2)})\widetilde t_1\widetilde t_2}
    \label{eq:16}
\end{equation}
This amplitude can be related to the ordinary one used in Quantum Mechanics as
\begin{eqnarray}
    -\frac{8\pi\sqrt{s}}{2M_N}(T^{\rm tot}_{K^*\bar K})^{-1} = (f^{QM})^{-1} \simeq -\frac{1}{a_0}+\frac{1}{2}r_0q_{cm}^2 - i q_{cm} 
    \label{eq:effectiverangeexpansion}
\end{eqnarray}
where $q_{cm}$ is the $p$ momentum in the rest frame of $pf_1(1285)$, from which we obtain the scattering length and effective range as 
\begin{eqnarray}
    \frac{1}{a_0} &=& \frac{8\pi\sqrt{s}}{2M_N} (T^{\rm tot}_{K^*\bar K})^{-1} \bigg|_{th} \nonumber \\
    r_0 &=& \frac{1}{\mu}\left[ \frac{\partial}{\partial\sqrt{s}} \left( -\frac{8\pi\sqrt{s}}{2M_N}(T^{\rm tot}_{K^*\bar K})^{-1}+iq_{cm} \right) \right]_{th}
    \label{eq:a0r0}
\end{eqnarray}
with $\mu$ being the $pf_1(1285)$ reduced mass. Elastic unitarity implies that the linear term in $q_{cm}$ in $\frac{8\pi\sqrt{s}}{2M_N} (T^{\rm tot}_{K^*\bar K})^{-1}$ is exactly $-iq_{cm}$, which is satisfied by $T^{\rm tot}_{K^*\bar K}$ of Eq.~(\ref{eq:16}) as shown analytically in \cite{Agatao:2025ckp}.

At this point, we must now consider the contribution of the second component of the $f_1$ wave function in Eq.~(\ref{eq:1}). Given that the two components of the wave function do not mix in the sequential $pf_1$ scattering steps, the two scattering amplitudes $T^{\rm tot}_i$ are evaluated independently for the two components. Elastic unitarity then demands that the total amplitude for the two components is obtained as 
\begin{eqnarray}
    (T^{\rm tot})^{-1} = \frac{1}{2} \left[ (T^{\rm tot}_{K^* \bar K})^{-1} + (T^{\rm tot}_{\bar K^* K})^{-1} \right]\, ,
    \label{eq:19}
\end{eqnarray}
where $T^{\rm tot}_{\bar K^* K}$ is obtained in the same manner from the $t_1'$ and $t_2'$ amplitudes of Eq.~\eqref{t_prime}.

\subsection{Correlation function}

For a single component of the cluster, the correlation function is defined as 
\begin{eqnarray}
    C_{pf_1}(p) = 1+4\pi &&\int_0^\infty dr\, r^2 S_{12}(r) \nonumber \\
    &&\times \Big\{ |j_0(pr)+TG|^2 - j_0^2(pr)\Big\},
\end{eqnarray}
where
\begin{eqnarray}
    TG = \Big(\widetilde T_{11}' + \widetilde T_{21}'\Big) G_1(\sqrt{s},r) + \Big(\widetilde T_{12}' + \widetilde T_{22}'\Big)G_2(\sqrt{s}, r)\, .
    \label{eq:21}
\end{eqnarray}
The source function $S_{12}(r)$ is defined as
\begin{eqnarray}
    S_{12}(r) = \frac{1}{(4\pi R^2)^{3/2}} e^{-r^2/4R^2}\, ,
\end{eqnarray}
where $R$ is the radius of the source, and the functions $G_i(\sqrt{s},r)$ in Eq.~(\ref{eq:21}) are given by
\begin{eqnarray}
    G_i(\sqrt{s},r) = \int && \frac{d^3 q}{(2\pi)^3} \frac{2M_N}{2E(q)2\omega_C(q)} \frac{F_C^{(i)}(q) j_0(qr)}{\sqrt{s}-E(q)-\omega_C(q)+i\epsilon} \nonumber \\
    &&\times \theta(q_{max}^{(i)} - q_i^*)\, .
\end{eqnarray}

Once again, we must take into account the two components of the $f_1(1285)$ wave function. In order to preserve the unitarity of the total scattering matrix, and given that $G_1(\sqrt{s},r),G_2(\sqrt{s},r)$ are very similar, we adopt the following unitary prescription for $TG$ Eq.~(\ref{eq:21}):
\begin{eqnarray}
    TG &\simeq& \Big( \widetilde T_{11}' + \widetilde T_{21}' + \widetilde T_{12}' + \widetilde T_{22}' \Big) \frac{1}{2}\Big( G_1(\sqrt{s},r) + G_2(\sqrt{s},r) \Big) \nonumber \\
    &=& T^{\rm tot}\frac{1}{2}\Big( G_1(\sqrt{s},r) + G_2(\sqrt{s},r) \Big) 
\end{eqnarray}
where $T^{\rm tot}$ is the average defined in Eq.~(\ref{eq:19}).

\section{Results}

\begin{figure}
    \centering
    \includegraphics[width=\linewidth]{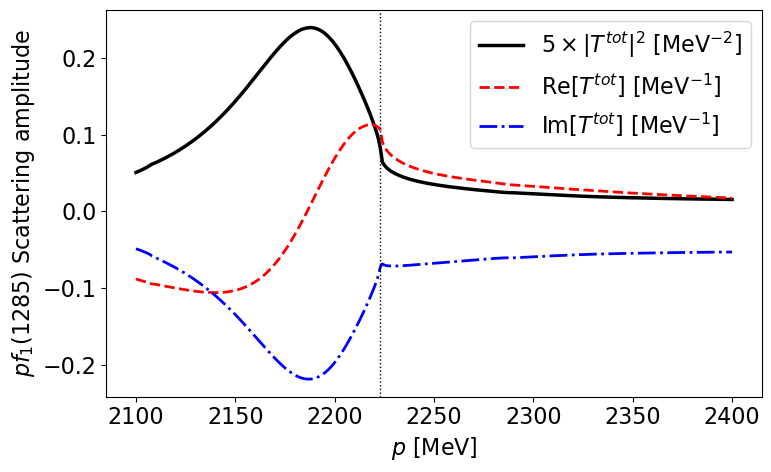}
    \caption{$pf_1(1285)$ scattering amplitude $T^{\rm tot}$ for the central values of the amplitude parameters, i.e.$g_{\Lambda(1800)}=3.65$, $M_{\Lambda(1800)}=1809$ MeV and $\Gamma_{\Lambda(1800)}=200$ MeV.}
    \label{fig:pf1_Tmatrix}
\end{figure}

We first show the results obtained for $T^{\rm tot}$ from Eq.~\eqref{eq:19} in Fig.~\ref{fig:pf1_Tmatrix}. The scattering matrix exhibits a resonant-like structure below threshold. The modulus of the imaginary part displays a pronounced peak around $2176$~MeV, while the real part changes sign from negative to positive at the peak of the imaginary part. The width of this state, estimated from ${\text Im}T^{\text tot}$, at half its maximum strength, is approximately $50$~MeV. This provides a clear prediction of the theoretical approach, indicating a three-body bound state roughly $35$~MeV below the $p~f_1(1285)$ threshold. The sub-threshold peak has a relatively large strength, which should facilitate its experimental observation. In practice, this could be studied by examining the invariant mass of a proton together with a decay product of the $f_1(1285)$, such as $a_0(980) \pi$.

The position and line shape of the scattering amplitude at the peak remain largely consistent with the results of~\cite{Encarnacion:2025lyf}, supporting the conclusion that the standard FCA, commonly used to predict three-body bound states, is a reliable tool for that purpose. The shapes of the real and imaginary parts of the amplitude around and above threshold are also very similar to those observed in Ref.~\cite{Encarnacion:2025lyf} (compare Fig.~\ref{fig:pf1_Tmatrix} here with Fig. 4 in Ref.~\cite{Encarnacion:2025lyf}). Yet, the strengths of the amplitudes along the studied energy range are somewhat different in the two approaches.

From this $T^{\rm tot}$ matrix, and using Eq.~\eqref{eq:a0r0} , we obtain the scattering length and effective range. The values obtained for the mass and width of the bound state, as well as the scattering parameters, are
$$ M_R = 2189\pm 10 \text{ MeV}\, , $$
$$ \Gamma_R = 53\pm 12 \text{ MeV} \, ,$$
$$ a_0 = [0.689\pm0.040] - [0.413\pm0.084]i \text{ fm} \, , $$
$$ r_0 = [-0.025\pm0.077] + [0.146\pm0.084]i \text{ fm} \, . $$
The central values and the uncertainties have been estimated by taking the parameter inputs $g_{\Lambda(1800)}=3.65\pm0.35$, $M_{\Lambda(1800)}=1809\pm9$ MeV and $\Gamma_{\Lambda(1800)}=200\pm50$ MeV. These parameters enter in the calculation of the amplitudes described in the Appendix of~\cite{Encarnacion:2025lyf}, in particular in the $p \bar K^*$ amplitude in $I=0$, which is assumed to be dominantly generated by the $\Lambda(1800)$ resonance. This contribution was parametrized using a Breit-Wigner form with the corresponding mass ($M_{\Lambda(1800)}$), width ($\Gamma_{\Lambda(1800)}$) and coupling ($g_{\Lambda(1800)}$). The associated uncertainties  were estimated by generating Gaussian-distributed random values of these parameters within the quoted uncertainties and evaluating the corresponding observables. The final uncertainty corresponds to the $68\%$ confidence level. 
 
 Note that this time there are changes with respect to Ref.~\cite{Encarnacion:2025lyf}. In that work we obtained $a_0= 1.04 -i 0.57$~fm and $r_0= 1.17 +i 1.16$~fm for the inputs $g_{\Lambda(1800)}=4$, $M_{\Lambda(1800)}=1800$~MeV, $\Gamma_{\Lambda(1800)}=200$~MeV\footnote{For a fair comparison, we provide the exact scattering parameters obtained with $g_{\Lambda(1800)}=4$, $M_{\Lambda(1800)}=1800$~MeV, $\Gamma_{\Lambda(1800)}=200$~MeV:
$a_0 = 0.638-0.374i \text{ fm}$ and 
$r_0 = 0.086+0.115i \text{ fm}$. Nevertheless, these scattering values fall well within the uncertainties of the present estimations at the $68\%$ confidence level, or close to its edge.}. The real part of the scattering length is now about one half of the value obtained in~\cite{Encarnacion:2025lyf}, while the imaginary part remains rather similar. On the other hand, the changes in the effective range are significant and we now obtain an effective range that is much smaller than that reported in~\cite{Encarnacion:2025lyf}.

\begin{figure}
    \centering
   \includegraphics[width=\linewidth]{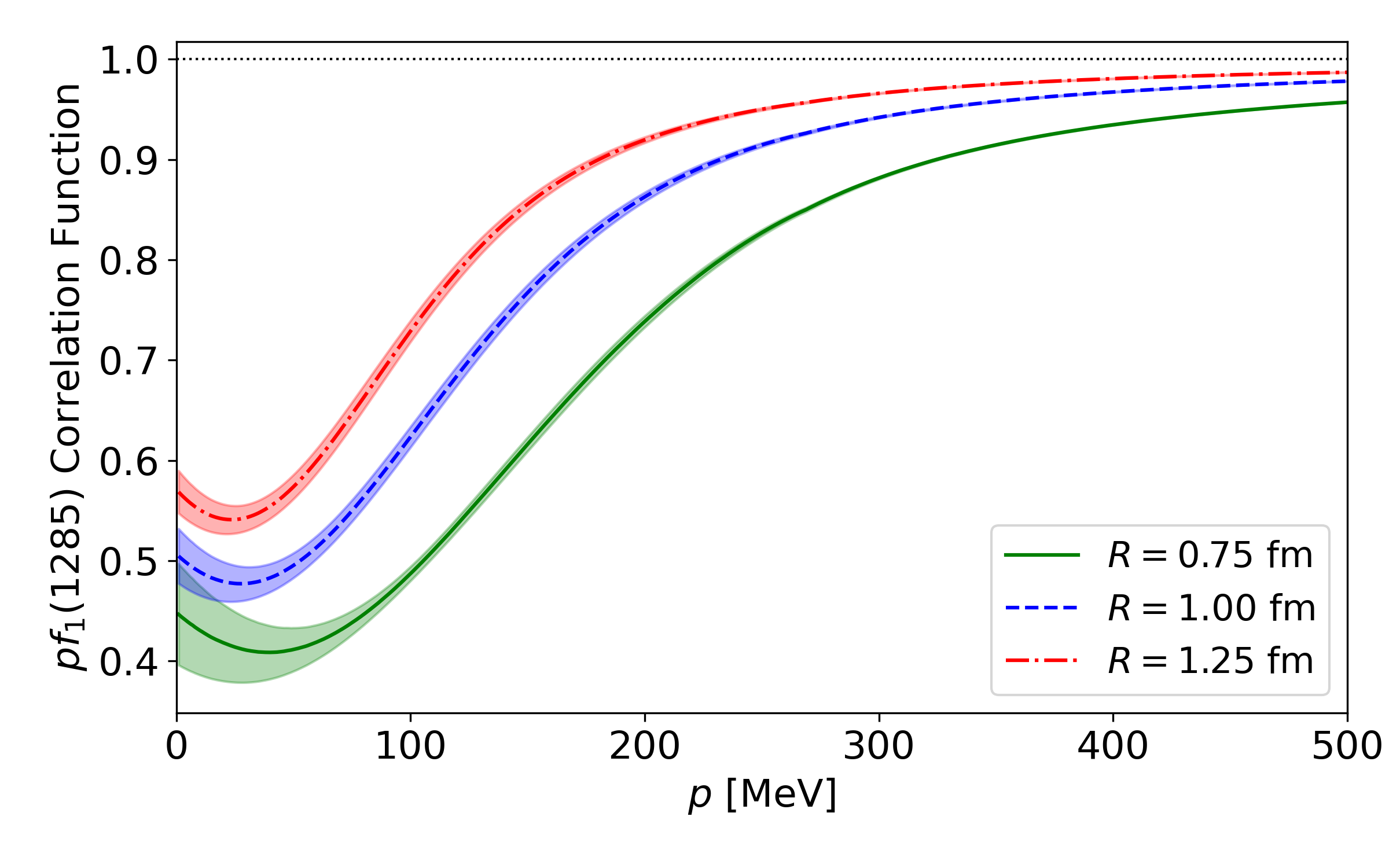}
   \caption{$pf_1(1285)$ CF with its corresponding error band estimation (see text for more details) for three different source size values.}
   \label{fig:pf1_CF}
\end{figure}

 Next we present the results obtained for the CF. Anticipating the experimental results from the ALICE collaboration, we perform the calculations for different values of the source size parameter of the CF, $R$. The results are shown in Fig.~\ref{fig:pf1_CF}. We display bands corresponding to different values of $R=0.75$, $1.00$ and $1.25$~fm. As mentioned above, we use the same amplitudes described in the Appendix of~\cite{Encarnacion:2025lyf} for the present calculations and estimate the error band in the same way by taking the theoretical uncertainties from the mass and width of the $\Lambda(1800)$ and from its coupling to $K \bar K^*$.

 The shapes of the CFs are similar, but their strengths differ slightly, changing from about $0.45$ at threshold for $R=0.75$~fm to about $0.55$ for $R=1.25$~fm. The results obtained are also qualitatively similar to those reported in Ref.~\cite{Encarnacion:2025lyf}. However, the new unitarization scheme adopted in this work, and the subsequent quantitative differences in the scattering amplitude with respect to that of Ref.~\cite{Encarnacion:2025lyf}, translate into a CF that approaches unity more slowly as a function of the momentum. Furthermore, the depletion at low momenta is shallower in the present case.
 
 We expect that the experimental results will be sufficiently precise to discriminate between these two behaviors, and indeed this is the case. In light of the recent preliminary experimental results presented at the Strangeness in Quark Matter 2026 conference~\cite{Serksnyte2026presentation}, only a few days after our prediction became publicly available, the results not only discriminate between these two behaviors but, most importantly, show very good agreement with our prediction for the $R=0.75$~fm case\footnote{Note that this preference for the source size is consistent with the experimental estimation of the effective source radius, which ranges from $0.69$ to $0.82$~fm.}. To illustrate the accuracy of the theoretical calculation in describing the $p$--$f_1(1285)$ CF, a comparative figure is provided in Appendix~\ref{sec:testing_prediction}.
  
\section {Conclusions}

We have revisited the work of Ref.~\cite{Encarnacion:2025lyf} on the interaction of the $p~f_1(1285)$ system in order to take into account recent developments that improve the FCA used in that work. In Ref.~\cite{Encarnacion:2025lyf}, a problem was identified in the FCA framework, namely that the resulting amplitude did not satisfy unitarity at the $p~f_1(1285)$ threshold. The unitarity was implemented there by rescaling the amplitude. In more recent works, the problem has been formally solved and a new algorithm has been provided that renders the amplitude unitary. Satisfying this condition is essential to reliably determine threshold scattering observables and CFs. 

In view of the imminent availability of experimental data for the $p~f_1$ CF, and to allow the ALICE collaboration to compare with the  most accurate results, we have recalculated all the results of Ref.~\cite{Encarnacion:2025lyf} using the new formalism. In retrospect, the new results show that the standard FCA framework already produces a peak below threshold at approximately the same energy and with a very similar shape.

  However, there are substantial differences for the scattering length and effective range of the $p~f_1$ system. The CF obtained with the new formalism  also changes, although the qualitative features remain similar. While the strength at threshold is comparable, the overall shape differs appreciably. 
  
The comparison of these predictions with the subsequent ALICE results, which show a reasonably good agreement, indicates consistency with the interpretation of the $f_1(1285)$ resonance as predominantly molecular, since this assumption underlies the theoretical calculation that reproduces the data with such accuracy. By extension, similar considerations may apply to other axial-vector resonances.

\begin{acknowledgments}
This work is partly supported by  the Spanish Ministerio de Economia y Competitividad (MINECO) and European FEDER funds under Contracts No. FIS2017-84038-C2-1-P B, PID2020- 112777GB-I00, and by Generalitat Valenciana under the contract PROMETEO/2020/023. This project has received funding from the European Union Horizon 2020 research and innovation program under the program H2020- INFRAIA2018-1, grant agreement No. 824093 of the STRONG-2020 project. This work is supported by the Spanish Ministerio de Ciencia e Innovación (MICINN) under contracts PID2020-112777GB-I00, PID2023-147458NB-C21 and CEX2023-001292-S; by Generalitat Valenciana under contracts PROMETEO/2020/023 and CIPROM/2023/59. P.\,E. and A.\,F. warmly thank the support from ACVJLI. %
\end{acknowledgments}

\appendix
\section{$p$~$f_1(1285)$ correlation function: comparison with experimental data}

\label{sec:testing_prediction}
\begin{figure}
    \centering
   \includegraphics[width=\linewidth]{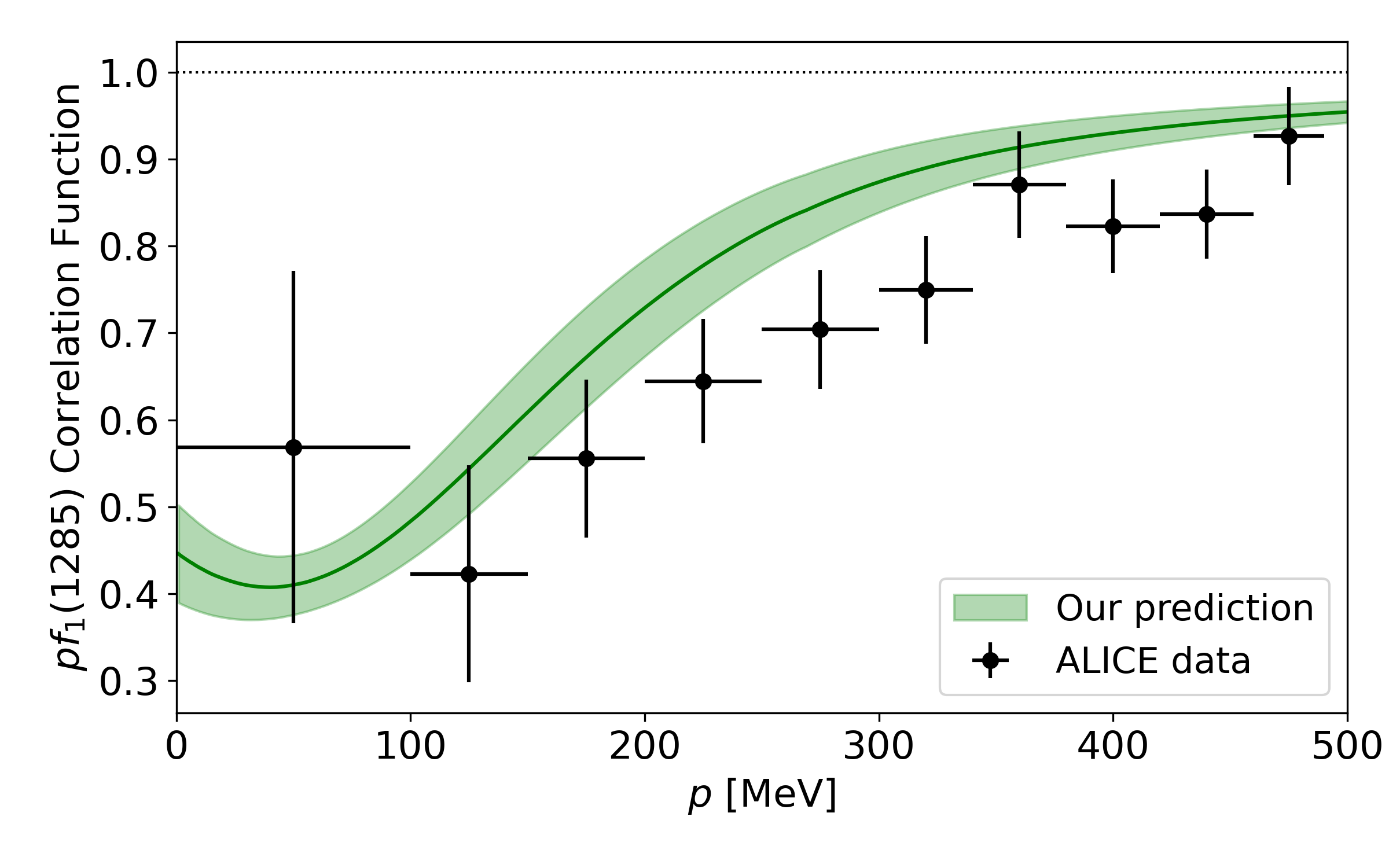}
   \caption{$pf_1(1285)$ CF for $R=0.75$~fm with the corresponding error band, compared with the ALICE experimental data taken from~\cite{Serksnyte2026presentation}.}
   \label{fig:pf1_CF_exp}
\end{figure}

In this section, we compare the theoretical calculation of the $p$--$f_1(1285)$ CF with the subsequent experimental measurement~\cite{Serksnyte2026presentation}. Figure~\ref{fig:pf1_CF_exp} confronts the theoretical prediction with the preliminary ALICE data. We keep only the calculation for $R=0.75$~fm of Fig.~\ref{fig:pf1_CF}, and take a range of $R$ between $0.69$ and $0.82$~fm for a fair comparison with the results of Ref.~\cite{Serksnyte2026presentation}. As can be seen in the figure, the predicted CF is in reasonable agreement with the experimental points. It is worth emphasizing that no fine-tuning of the theoretical model parameters was performed.

\bibliography{apssamp.bib}

\end{document}